\documentclass[aps,prd,nofootinbib,amsmath,amssymb,superscriptaddress,twocolumn,10pt]{revtex4}%superscriptaddress,showpacs,
\usepackage{graphicx}
\usepackage{dcolumn}
\usepackage{bm}
\usepackage{amssymb}
\usepackage{latexsym}
\usepackage{booktabs}
\usepackage{amsmath}
\usepackage{multirow}
\usepackage[colorlinks=true, linkcolor=red, citecolor=blue]{hyperref}

\newcommand{\be}{\begin{equation}}
\newcommand{\ee}{\end{equation}}
\newcommand{\bq}{\begin{eqnarray}}
\newcommand{\eq}{\end{eqnarray}}

\begin{document}

\title{Probing $f(R)$ cosmology with sterile neutrinos via measurements of scale-dependent growth rate of structure}

\author{Yun-He Li}
\affiliation{Department of Physics, College of Sciences, Northeastern University, Shenyang
110004, China}
\author{Jing-Fei Zhang}
\affiliation{Department of Physics, College of Sciences, Northeastern University, Shenyang
110004, China}
\author{Xin Zhang\footnote{Corresponding author}}
\email{zhangxin@mail.neu.edu.cn} \affiliation{Department of Physics, College of Sciences,
Northeastern University, Shenyang 110004, China}
\affiliation{Center for High Energy Physics, Peking University, Beijing 100080, China}

\begin{abstract}
In this paper, we constrain the dimensionless Compton wavelength parameter $B_0$ of $f(R)$ gravity as well as the mass of sterile neutrino by using the cosmic microwave background observations, the baryon acoustic oscillation surveys, and the linear growth rate measurements. Since both the $f(R)$ model and the sterile neutrino generally predict scale-dependent growth rates, we utilize the growth rate data measured in different wavenumber bins with the theoretical growth rate approximatively scale-independent in each bin. The employed growth rate data come from the peculiar velocity measurements at $z=0$ in five wavenumber bins, and the redshift space distortions measurements at $z=0.25$ and $z=0.37$ in one wavenumber bin. By constraining the $f(R)$ model alone, we get a tight 95\% upper limit of $\log_{10}B_0<-4.1$. This result is slightly weakened to $\log_{10}B_0<-3.8$ (at 2$\sigma$ level) once we simultaneously constrain the $f(R)$ model and the sterile neutrino mass, due to the degeneracy between the parameters of the two. For the massive sterile neutrino parameters, we get the effective sterile neutrino mass $m_{\nu,{\rm{sterile}}}^{\rm{eff}}<0.62$ eV (2$\sigma$) and the effective number of relativistic species $N_{\rm eff}<3.90$ (2$\sigma$) in the $f(R)$ model. As a comparison, we also obtain $m_{\nu,{\rm{sterile}}}^{\rm{eff}}<0.56$ eV (2$\sigma$) and $N_{\rm eff}<3.92$ (2$\sigma$) in the standard $\Lambda$CDM model.

\end{abstract}

 \maketitle

%\section{Introduction}

%The acceleration of the current universe's expansion leaves an open question of what mysterious force is responsible for this phenomenon \cite{Riess98}.

One of the most profound puzzles in contemporary physics is the cause for the current cosmic acceleration \cite{Riess98,Perlmutter98}.
Although the mainstream for explaining this late-time cosmic acceleration is by introducing an extra dark energy fluid under the framework of general relativity (a typical dark energy candidate is cosmological constant $\Lambda$, corresponding to a $\Lambda$CDM cosmology) \cite{de1,de2,de3}, other possibility like the modification to general relativity on large scales also attracts more and more attention. A simple modified gravity scenario is the so-called $f(R)$ theory which generates acceleration via adding a function of the Ricci scalar $R$ to the Einstein-Hilbert action \cite{fR1,fR2}. For recent reviews of $f(R)$ theory, see, e.g., Refs.~\cite{Sotiriou:2008rp,DeFelice:2010aj,Clifton:2011jh,Weinberg:2012es,Joyce:2014kja}.

Generally, $f(R)$ theory introduces an extra scalar degree of freedom $f_R\equiv df/dR$ characterized by a squared Compton wavelength proportional to $f_{RR}\equiv d^2f/dR^2$, or by a dimensionless quantity \cite{B}
\begin{equation}\label{Bdefinition}
    B\equiv {f_{RR}\over1+f_R}{dR\over d\ln a}\left({d\ln H\over d\ln a}\right)^{-1},
\end{equation}
where $H=\dot{a}/a$ is the Hubble parameter, $a$ is the scale factor, and a dot denotes the derivative with respect to the cosmic time $t$. Due to the existence of this extra scalar degree of freedom, $f(R)$ models tend to enhance the growth of the matter perturbations below the Compton wavelength scale. Thus $f(R)$ models generally predict higher and scale-dependent growth rates compared with the standard $\Lambda$CDM model. By accurately measuring the growth rate of large-scale structure, we therefore have a powerful tool to constrain $f(R)$ models.

A direct way of measuring the large-scale growth rate comes from redshift-space distortions (RSD) in galaxy surveys \cite{RSD1,RSD2}, which depict the distortions of the measured galaxy maps in redshift space. Such distortions arise from the peculiar velocities of galaxies, which actually relate to the evolution of matter perturbation $\delta_m$. Thus RSD provide information about the large-scale growth rate $f_m=d\ln\delta_m/d\ln a$ or the bias-independent growth rate $f_m\sigma_8(z)$ with $\sigma_8(z)$ the root-mean-square mass fluctuation in spheres with radius $8h^{-1}$ Mpc at redshift $z$.\footnote{Note that, in practice, $\sigma_8(z)$ is computed in linear regime. Although the expression, $\sigma_8^2(z)=\frac{1}{2\pi^2}\int_0^{\infty}dk k^2P(k,z)W^2(kr)$, where $r=8h^{-1}\,{\rm Mpc}$ and $P(k,z)$ is the matter power spectrum, contains an integral on all scales, the top-hat window function, $W(kr)=3(\sin{kr}-kr\cos{kr})/(kr)^3$, decays rapidly when $k$ is beyond the linear regime, and thus the nonlinear modes contribute very little to the integral.} Recent galaxy surveys have measured $f_m\sigma_8(z)$ in different scale ranges with precision up to 7\% \cite{RSD6dF,RSD2dF,RSDwigglez,RSDsdss7,Beutler:2013yhm,RSDvipers}. Some works also have used these data to constrain some specific $f(R)$ models \cite{Okada:2012mn,Xu:2014wda}.

\begin{figure}[tbp]
\includegraphics[width=7cm]{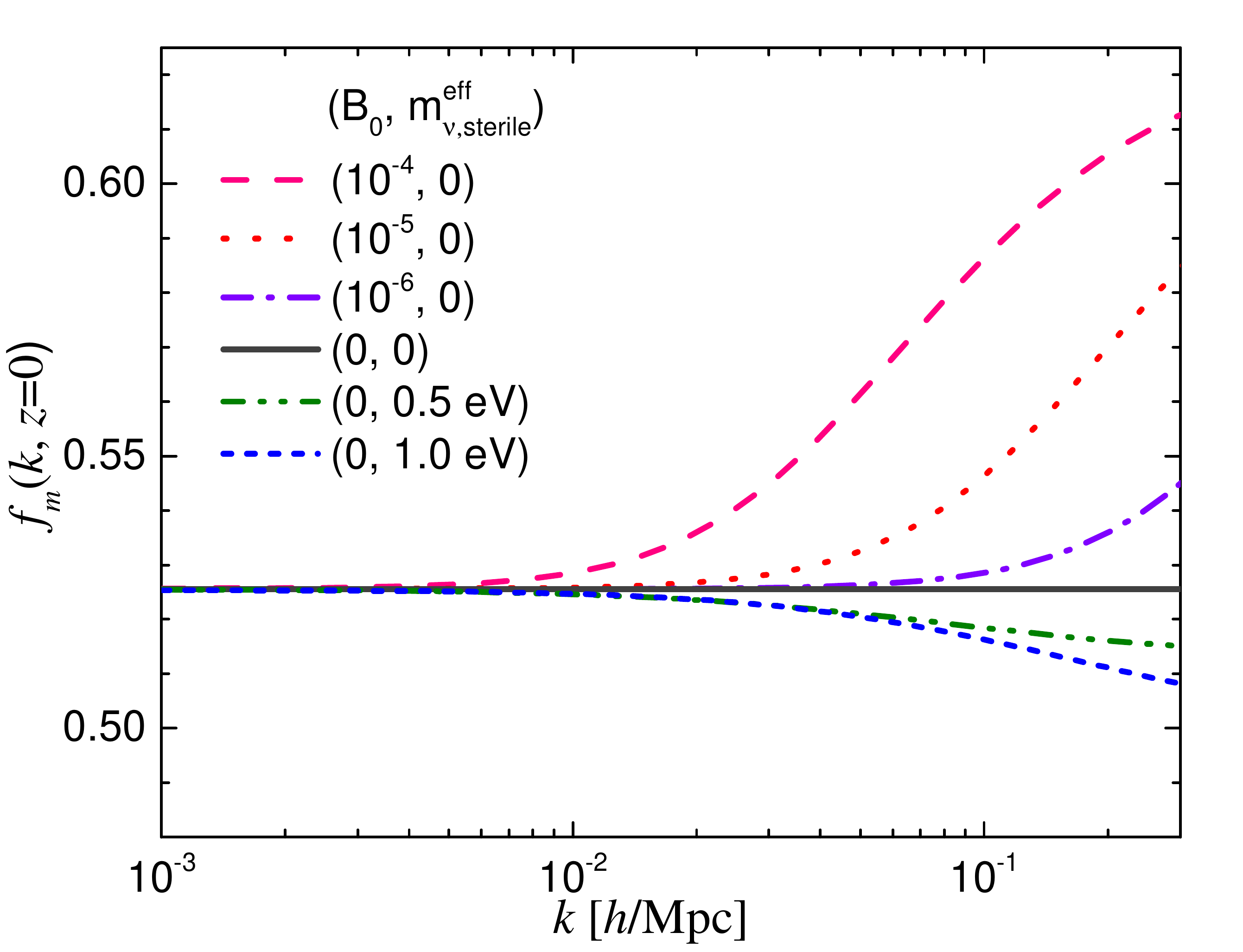}
\caption{\label{fig1} The scale-dependent linear growth rate $f_m(k,z)$ at $z=0$ for $B_0=10^{-4}$ (pink dashed), $10^{-5}$ (red dot), and $10^{-6}$ (violet dashed dot) of $f(R)$ model, and $m_{\nu,{\rm{sterile}}}^{\rm{eff}}=0.5$ eV (olive dashed dot dot) and $1.0$ eV (blue short dashed) of sterile neutrino. The black solid line denotes the growth rate for the standard $\Lambda$CDM model. Note that we fix $\Omega_m$ at the same values for all the curves, and $N_{\rm eff}=4.046$ for the curves of sterile neutrino. It is clear that $\Lambda$CDM model gives a scale-independent growth rate, while both $f(R)$ and sterile neutrino predict scale-dependent growth rates. Besides, a larger $B_0$ tends to enhance the growth rate more, while $m_{\nu,{\rm{sterile}}}^{\rm{eff}}$ affects the growth rate in the opposite way.}
\end{figure}

Nevertheless, current RSD models used to extract $f_m$ from the power spectrum of galaxy survey always assume a scale-independent $f_m$ on all scales, which is true for the standard $\Lambda$CDM model but not valid for the $f(R)$ cosmology. To see clearly, we plot the general $k$-dependent growth rate $f_m(k,\,z)$ at $z=0$ in Fig.~\ref{fig1} for the standard $\Lambda$CDM and $f(R)$ models with $B_0=10^{-4}$, $10^{-5}$ and $10^{-6}$, where $k$ is the comoving wavenumber and the subscript ``0'' denotes the current value of the corresponding quantity. The plotting range for $k$ covers most of current RSD surveys' scales. It is clear that the growth rate in the $f(R)$ model significantly deviates from the scale-independence. Particularly for the $B_0=10^{-4}$ case, the deviation arrives about 14\% at $k=0.3h\,\rm{Mpc^{-1}}$, even larger than current RSD surveys' precisions. So it will induce substantial systematic errors for the fit results when we simply use all the RSD data to constrain the $f(R)$ models without careful selection according to their measured scales. Actually, by performing N-body simulations, Ref.~\cite{Jennings:2012pt} illustrated that the measured growth rate from RSD surveys agrees with that predicted in $f(R)$ model only for $k<0.06h\,\rm{Mpc^{-1}}$ at $z=0$.

%=======================Figure 1======================================
%=====================================================================

To avoid such systematic errors, we constrain $f(R)$ model using the scale-dependent growth rate data in this work. Recently, Ref.~\cite{Johnson:2014kaa} tried to measure the scale-dependent growth rate by using the observations of peculiar velocities of galaxies from 6dF Galaxy Survey velocity sample in combination with a newly-compiled sample of low-redshift type Ia supernovae. The measurement obtained the scale-dependent growth rates in five $k$ bins at $z=0$: $f_m\sigma_{8}(k)=0.79\pm0.21,\,0.30\pm0.14,\,0.32\pm0.19,\,0.64\pm0.17$, and $0.48\pm0.22$, for $k \in[0.005,\,0.02]$, $[0.02,\,0.05]$, $[0.05,\,0.08]$, $[0.08,\,0.12]$, and $[0.12,\,0.15] h\,\rm{Mpc^{-1}}$, respectively.\footnote{The original data are not gaussian. For the purpose of using these data to constrain $f(R)$ model, we choose the upper limits of the data as the global errors. The growth rates are measured in a series of $\Delta k\sim0.03h\,{\rm Mpc^{-1}}$ bins in Ref.~\cite{Johnson:2014kaa}. We assume that each data point is measured independently. We calculate the theoretical value of $f_m\sigma_8(k_i)$ in each bin at $k_i=k_i^{\rm cen}$ with $k_i^{\rm cen}$ the centre of the $i$th bin, and then compare it with the measurement value of the $i$th bin.} Hereafter, we use PV to denote these five data points. Besides, we also use the growth rate measurements $f_m\sigma_{8}(z)=0.35\pm0.06$ at $z=0.25$ and $f_m\sigma_{8}(z)=0.46\pm0.04$ at $z=0.37$ from SDSS DR7 \cite{RSDsdss7}. These two data points actually belong to the RSD measurement but are obtained from the power spectrum with $k\in[0.005,\,0.033]h\,\rm{Mpc^{-1}}$. In this scale range, the growth rate in $f(R)$ model can be approximatively considered scale-independent from Fig.~\ref{fig1}. So it is appropriate to use these two data to constrain the $f(R)$ model. Hereafter, we use RSD to denote these two data points.

A successful $f(R)$ model should give a stable acceleration solution at late times and also pass the local solar system tests. Several viable $f(R)$ models that satisfy these conditions have been proposed in the literature \cite{fRmodel1,fRmodel2,fRmodel3,fRmodel4,fRmodel5,fRmodel6,fRmodel7,fRmodel8,fRmodel9}. However, we only care about the large-scale perturbations of $f(R)$ models in this paper. So we consider a parametrized $f(R)$ model whose background is the same as that of $\Lambda$CDM and perturbations are parametrized by two functions \cite{Bertschinger:2008zb,Giannantonio:2009gi},
\begin{eqnarray}\label{BZ}
&&\mu(k,a)=\frac{1}{1-1.4 \cdot 10^{-8}|\lambda_1|^2a^3}\frac{1+\frac{4}{3}\lambda_1^2\,k^2a^4}{1+\lambda_1^2\,k^2a^4}, \nonumber \\
&&\gamma(k,a)=\frac{1+\frac{2}{3}\lambda_1^2\,k^2a^4}{1+\frac{4}{3}\lambda_1^2\,k^2a^4},
\end{eqnarray}
where the parameter $\lambda_1$ has dimension of length, which can be related to the current value of $B$ via $\lambda_1^2=B_0\,c^2/(2H_0^2)$ with $c$ the speed of light. The functions $\mu(k,a)$ and $\gamma(k,a)$ are used in the public MGCAMB code \cite{Lewis:1999bs,Hojjati:2011ix} to quantify the modifications to the Poisson and anisotropy equations:
\begin{eqnarray}
&&k^2\Psi = - \mu(k,a) 4 \pi G a^2 \lbrace \rho \Delta + 3(\rho + P) \sigma \rbrace  \label{mg-poisson}, \\
&&k^2[\Phi - \gamma(k,a) \Psi] = \mu(k,a)  12 \pi G a^2   (\rho + P) \sigma \label{mg-anisotropy},
\end{eqnarray}
where $\Psi$ and $\Phi$ are two metric perturbation potentials, $\Delta$ is the comoving density perturbation, and $\sigma$ denotes the anisotropic stress. It was showed that Eq.~(\ref{BZ}) together with Eqs.~(\ref{mg-poisson}) and (\ref{mg-anisotropy}) can well model the large scale growth in $f(R)$ models for $B_0\lesssim1$ \cite{Giannantonio:2009gi}. In our paper, we will constrain $B_0$ using above-mentioned growth rate data. Since the growth rate is very sensitive to the order of magnitude of $B_0$, we will take $\log_{10}B_0$ as a free parameter.

Besides the value of $B_0$ of $f(R)$ model, another interest of this work is the mass of massive neutrinos. The total mass of massive active neutrinos in a $f(R)$ universe has been widely reported in previous works, e.g., Refs.~\cite{Hojjati:2011ix,He:2013qha,Dossett:2014oia,Zhou:2014fva,Geng:2014yoa}. So we here focus on the mass of massive sterile neutrinos. Recently, the extra massive sterile neutrino species has attracted much attention because of its magic power of simultaneously alleviating the tensions between different observations \cite{sterile1,sterile2,sterile3,sterile4,sterile5,sterile6,Zhang:2014nta,Li:2014dja,Zhang:2014ifa,Zhang:2014lfa}. To describe the mass of sterile neutrino, we need two extra free parameters, the effective sterile neutrino mass $m_{\nu,{\rm sterile}}^{\rm eff}$ and the effective number of relativistic species $N_{\rm eff}$. With these two parameters, the true mass of a thermally-distributed sterile neutrino reads $m_{\rm sterile}^{\rm thermal}=(N_{\rm eff}-3.046)^{-3/4}m_{\nu,{\rm sterile}}^{\rm eff}$. In order to avoid a negative $m_{\rm sterile}^{\rm thermal}$, $N_{\rm eff}$ must be larger than 3.046 in a universe with sterile neutrinos. The mass of sterile neutrino also corresponds to a characteristic scale, below which sterile neutrino is free-streaming and tends to suppress the growth of structure. Thus sterile neutrino can also lead to a scale-dependent growth rate (see Fig.~\ref{fig1}). There might exist a degeneracy between $B_0$ and the mass of sterile neutrino when we use the growth rate data to simultaneously constrain the two. We will constrain the parameters of sterile neutrino in the $f(R)$ model and verify whether this degeneracy exists or not.

%===================Table 1===========================================
\begin{table*}[tbp]
\centering\caption{\label{table1} The mean values and 1$\sigma$ errors for the main parameters of the $f(R)$ model, the $\Lambda$CDM+$m_{\nu,{\rm{sterile}}}^{\rm{eff}}+N_{\rm eff}$ model, and the $f(R)$+$m_{\nu,{\rm{sterile}}}^{\rm{eff}}+N_{\rm eff}$ model from the CMB+BAO+PV and CMB+BAO+PV+RSD data combinations. For the parameters $m_{\nu,{\rm{sterile}}}^{\rm{eff}}$, $N_{\rm eff}$, and $\log_{10}B_0$, we show their 95\% upper limits and mean values in brackets. Note that the mass of sterile neutrino $m_{\nu,{\rm{sterile}}}^{\rm{eff}}$
is in unit of eV and the Hubble constant $H_0$ is in unit of km s$^{-1}$ Mpc$^{-1}$.}
\begin{tabular}{lcccccccc}
\hline
\hline&\multicolumn{2}{c}{$f(R)$ }&&\multicolumn{2}{c}{$\Lambda$CDM+$m_{\nu,{\rm{sterile}}}^{\rm{eff}}+N_{\rm eff}$}&&\multicolumn{2}{c}{$f(R)$+$m_{\nu,{\rm{sterile}}}^{\rm{eff}}+N_{\rm eff}$}\\
           \cline{2-3}\cline{5-6}\cline{8-9}
Parameter & CMB+BAO+PV&+RSD&&CMB+BAO+PV&+RSD&&CMB+BAO+PV&+RSD \\
\hline
$m_{\nu,{\rm{sterile}}}^{\rm{eff}}$&...&...&&$<0.43(0.18)$&$<0.56(0.23)$&&$<0.61(0.22)$&$<0.62(0.24)$\\
$N_{\rm eff}$&...&...&&$<3.96(3.47)$&$<3.92(3.43)$&&$<3.95(3.45)$&$<3.90(3.40)$\\
$\log_{10}B_0$&$<-2.7(-6.1)$&$<-4.1(-7.0)$&&...&...&&$<-1.8(-5.6)$&$<-3.8(-6.8)$\\
$\Omega_m$&$0.3035\pm0.0071$&$0.3027^{+0.0069}_{-0.0075}$&&$0.3047\pm0.0080$&$0.3047^{+0.0078}_{-0.0079}$&&$0.3040^{+0.0080}_{-0.0081}$&$0.3047^{+0.0078}_{-0.0084}$\\
$\sigma_8$&$0.850^{+0.006}_{-0.049}$&$0.827^{+0.008}_{-0.018}$&&$0.809^{+0.029}_{-0.022}$&$0.799^{+0.031}_{-0.023}$&&$0.851^{+0.032}_{-0.079}$&$0.807^{+0.031}_{-0.030}$\\
$H_0$&$68.1\pm0.6$&$68.2\pm0.6$&&$69.8^{+1.1}_{-1.8}$&$69.6^{+1.0}_{-1.7}$&&$69.8^{+1.1}_{-1.8}$&$69.5^{+1.0}_{-1.7}$\\
\hline
$-\ln\mathcal{L}_{\rm{max}}$ &4912.405&4913.993&&4912.085 &4913.852 && 4912.361&4914.293 \\
\hline
\end{tabular}
\end{table*}

%In the following, we will present the main constraint results.
Our calculation is based on the public Markov-Chain Monte-Carlo package CosmoMC \cite{Lewis:2002ah}. The free parameter vector is: $\{\omega_b,\, \omega_c,\, \theta_{\rm{MC}},\, \tau,\,n_{\rm{s}},\, {\rm{ln}}(10^{10}A_{\rm{s}}),\,\log_{10}B_0,\,m_{\nu,{\rm sterile}}^{\rm eff},\,N_{\rm eff}\}$, where $\omega_b$, $\omega_c$, and $H_0$ denote the background physical baryon density, physical cold dark matter density, and Hubble constant today, respectively, $\theta_{\rm{MC}}$ is approximation to the angular size of the sound
horizon at the time of last-scattering, $\tau$ denotes the optical depth to reionization, and ${\rm{ln}}(10^{10}A_{\rm{s}})$ and $n_{\rm{s}}$ are the amplitude and the spectral index of the primordial scalar perturbation power spectrum. Following Ref.~\cite{Dossett:2014oia}, we set a prior [$-10$, 1] for $\log_{10}B_0$. The priors of other free parameters are the same as those used by Planck Collaboration \cite{Ade:2013zuv}. For massive active neutrino species, we assume its total mass $\sum m_{\nu}=0.06$ eV.

To constrain other cosmological parameters and break degeneracies between them, we also employ the observational data from cosmic microwave background (CMB) and baryon acoustic oscillation (BAO) observations. For the CMB data, we use the temperature power spectrum data $C_\ell^{TT}$ \cite{Ade:2013zuv} and lensing data $C_\ell^{\phi\phi}$~\cite{Ade:2013tyw} from Planck\footnote{After the completion of this work, the Planck 2015 results were reported (on arXiv). Note also that currently the Planck 2015 likelihoods are not yet available. So the Planck data used in this work come from the Planck 2013 release.} in combination with the WMAP 9-yr polarization (TE and EE) power spectrum data \cite{wmap9}. For the BAO data, we use the measurements from 6dFGS ($z=0.1$) \cite{6df}, SDSS DR7 ($z=0.35$) \cite{sdss7}, WiggleZ ($z=0.44$, 0.60, and 0.73) \cite{wigglez}, and BOSS DR11 ($z=0.32$ and 0.57) \cite{boss}. Note that the late-time integrated Sachs-Wolfe (ISW) effect and the lensing signal from the CMB measurements can also give constraint on $B_0$, since $B_0$ affects the evolutions of the metric perturbation potentials $\Phi$ and $\Psi$ from Eqs.~(\ref{BZ})--(\ref{mg-anisotropy}), and the ISW effect and CMB lensing potential are directly related to $\dot{\Phi}+\dot{\Psi}$ and $\Phi+\Psi$, respectively.

%\section{Results and discussions}

%\subsection{Growth index and sterile neutrinos}

%========================Figure 2======================================
\begin{figure}[tbp]
\centering % \begin{center}/\end{center} takes some additional vertical space
\includegraphics[width=7cm]{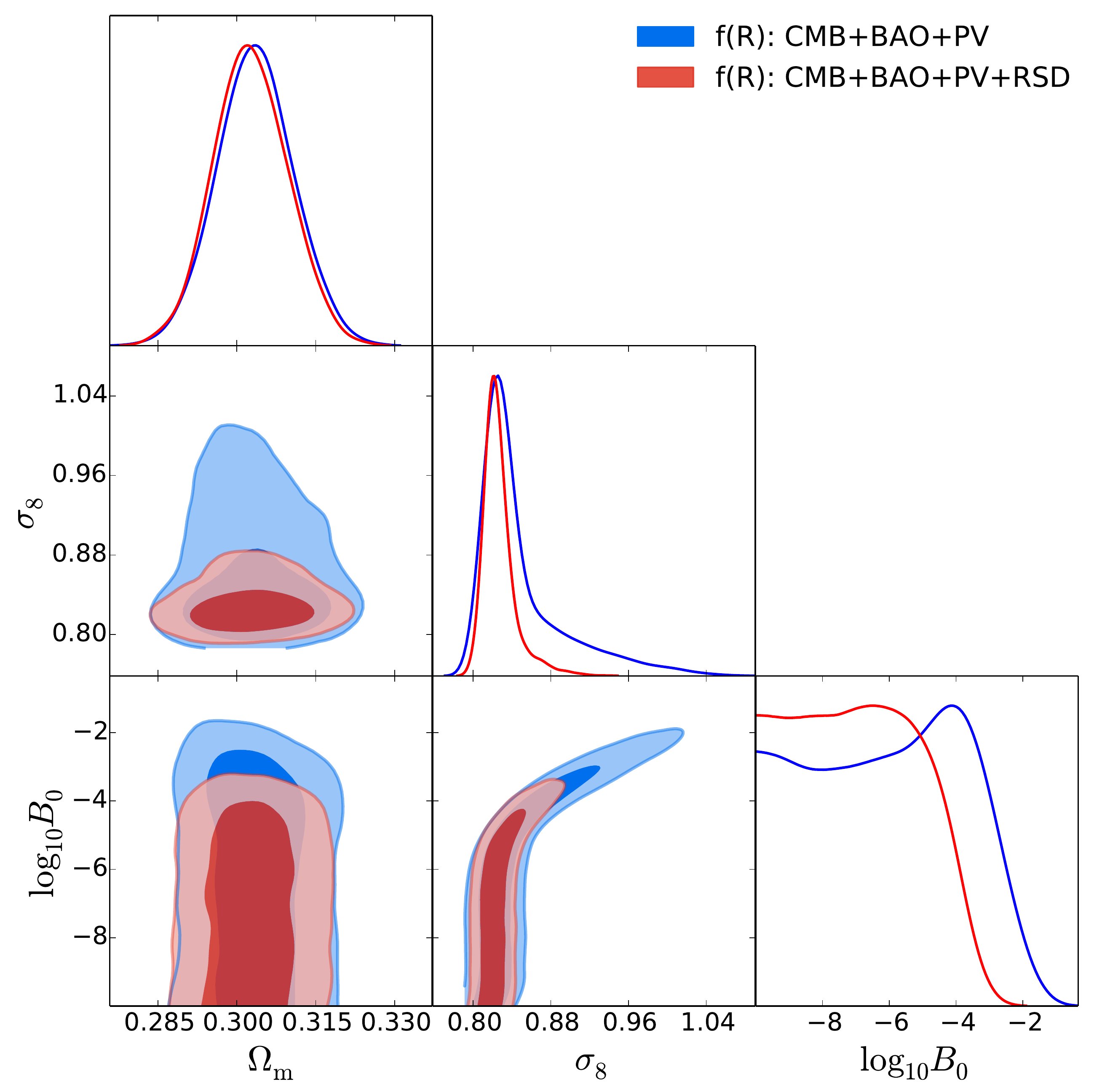}
\hfill
%\includegraphics[width=.45\textwidth,origin=c,angle=180]{img2.pdf}
% "\includegraphics" is very powerful; the graphicx package is already loaded
\caption{\label{fig2} The one-dimensional posterior distributions and two-dimensional marginalized contours (68\% and 95\% CL) for the $f(R)$ model from the CMB+BAO+PV and CMB+BAO+PV+RSD data combinations.}
\end{figure}
%====================================================================

First, we constrain the $f(R)$ parameter $B_0$ without considering the extra sterile neutrino species. The main fit results are shown in Table \ref{table1} and Fig. \ref{fig2}. By using the CMB+BAO data and the scale-dependent growth rate PV data, we obtain the 95\% upper limit of $\log_{10}B_0<-2.7$. Here note that the main contribution to the constraint result of $B_0$ comes from the PV data, since current CMB+BAO data (with the CMB lensing measurement) can only give a weak constraint on $B_0$, e.g., $B_0<0.12$ at 2$\sigma$ level \cite{Hu:2013aqa}. From Fig. \ref{fig2}, we can see a weak peak for $\log_{10}B_0$ at about $-4$ from the CMB+BAO+PV data. A peak with a nonzero $B_0$ means that a relatively large growth rate is more favored by the data. The PV data in the first and fourth $k$ bins may dominate the appearance of this peak, since the growth rates measured in these two bins are noticeably large. However, this peak for $\log_{10}B_0$ disappears once we further add the RSD data into our analysis. Besides, the 95\% upper limit of $\log_{10}B_0$ is improved to $\log_{10}B_0<-4.1$ with the help of the RSD data. These changes result from the fact that the RSD data favor a lower growth rate and have higher precision than the PV data.

Our fit result for $B_0$ from the CMB+BAO+PV+RSD data is comparable to the limit, $\log_{10}B_0<-4.07$ (at 2$\sigma$ level), quoted in Ref.~\cite{Dossett:2014oia}, where the WiggleZ galaxy power spectrum with $k_{\rm max}=0.2h\,{\rm Mpc}^{-1}$ is used, and is also comparable to the limit from the recent Planck 2015 result, $\log_{10}B_0<-4.01$ (at 2$\sigma$ level), obtained by combining the RSD data at $z=0.57$ \cite{Ade:2015rim}. These results represent the overall levels of constraint on the parametrized $f(R)$ model that can be achieved from the current information on linear scales. A tighter constraint on $f(R)$ model can be obtained if the probes on smaller scales are utilized, such as the Planck Sunyaev-Zeldovich (PSZ) cluster counts \cite{Boubekeur:2014uaa}, the strong lensing galaxy signals \cite{Smith:2009fn}, the distance indicators \cite{Jain:2012tn}, and the fifth-force effects on diffuse dwarf galaxy components \cite{Vikram:2013uba,Vikram:2014uza}. However, it is more difficult to control the possible systematic errors of these small-scale measurements. For example, the tension between the Planck temperature power spectrum measurement and the PSZ cluster counts may be a hint of some unknown systematic errors existing in the PSZ measurement. Besides, the physics of $f(R)$ theory on nonlinear scales is also complicated. So we do not incorporate the information on small scales in our analysis.

\begin{figure}[tbp]
\centering % \begin{center}/\end{center} takes some additional vertical space
\includegraphics[width=7cm]{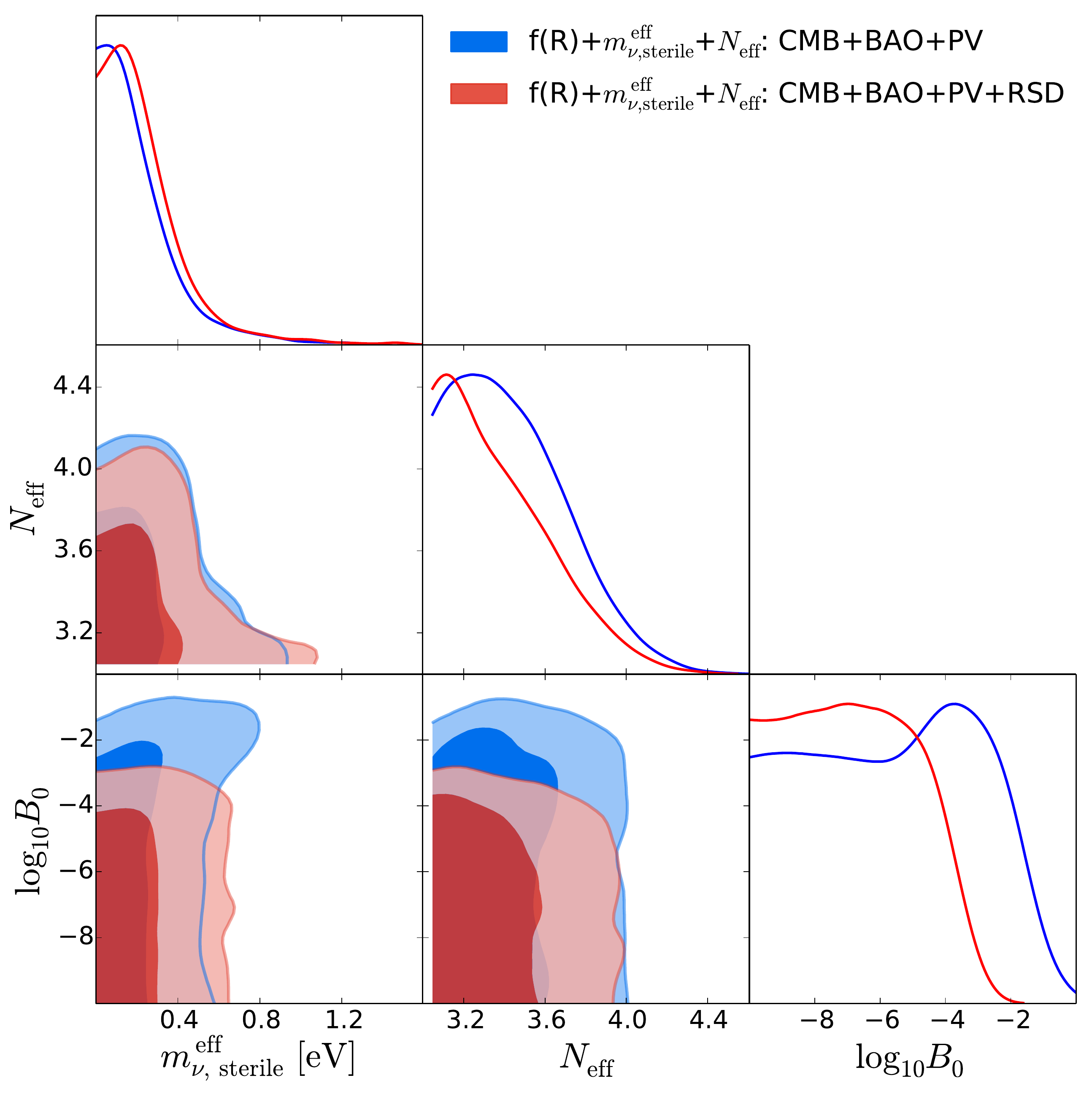}
\hfill
%\includegraphics[width=.45\textwidth,origin=c,angle=180]{img2.pdf}
% "\includegraphics" is very powerful; the graphicx package is already loaded
\caption{\label{fig3} The one-dimensional posterior distributions and two-dimensional marginalized contours (68\% and 95\% CL) for the $f(R)$+$m_{\nu,{\rm{sterile}}}^{\rm{eff}}+N_{\rm eff}$ model from the CMB+BAO+PV and CMB+BAO+PV+RSD data combinations.}
\end{figure}

Next, we consider the cosmological constraints on $f(R)$ parameter in a universe with sterile neutrinos. The fit results from the CMB+BAO+PV and CMB+BAO+PV+RSD data combinations are shown in Table \ref{table1} and Fig. \ref{fig3}. From Fig. \ref{fig3}, we also see a peak for $\log_{10}B_0$ at about $-4$ for the CMB+BAO+PV data. Comparing Fig. \ref{fig3} with Fig. \ref{fig2}, we find that there indeed exists a slight degeneracy between the $f(R)$ parameter and the sterile neutrino parameters. Due to this degeneracy, the 95\% upper limits of $\log_{10}B_0$ are enlarged to $\log_{10}B_0<-1.8$ and $<-3.8$ for the CMB+BAO+PV data and the CMB+BAO+PV+RSD data, respectively. For the sterile neutrino parameters, we get $m_{\nu,{\rm{sterile}}}^{\rm{eff}}<0.61$ eV and $N_{\rm eff}<3.95$ (2$\sigma$) from the CMB+BAO+PV data, and $m_{\nu,{\rm{sterile}}}^{\rm{eff}}<0.62$ eV and $N_{\rm eff}<3.90$ (2$\sigma$) from the CMB+BAO+PV+RSD data. To see how the $f(R)$ gravity affects the fit value of the sterile neutrino mass, we also constrain the sterile neutrino parameters in the $\Lambda$CDM universe by using the same observational data. The main fit results are also shown in Table \ref{table1}. The CMB+BAO+PV data give $m_{\nu,{\rm{sterile}}}^{\rm{eff}}<0.43$ eV and $N_{\rm eff}<3.96$ (2$\sigma$) in the $\Lambda$CDM universe. When the RSD data are also taken into account, the constraint results become $m_{\nu,{\rm{sterile}}}^{\rm{eff}}<0.56$ eV and $N_{\rm eff}<3.92$ (2$\sigma$). We can find that the 95\% upper limits of the effective sterile neutrino mass are also enlarged in the $f(R)$ gravity, which also results from the degeneracy between the $f(R)$ parameter and the sterile neutrino parameters.

%\subsection{With tensor modes and BICEP2 data}

%=======================Figure 3======================================
%===================================================================

%\section{Conclusion}

Current cosmic acceleration can result from either a dark energy fluid or a modification to gravity on large scales. Among various modified gravity theories, the so-called $f(R)$ theory is a simple one. The key feature of the $f(R)$ theory that differs from the standard $\Lambda$CDM model is that $f(R)$ models generally predict higher and scale-dependent growth rates. In this paper, we focus on a parametrized $f(R)$ model, whose power of modification to the large-scale structure growth is described by a dimensionless Compton wavelength parameter $B_0$. We constrain $\log_{10}B_0$ by using the CMB data, the BAO data, and the linear growth rate data. To avoid the possible systematic errors when constraining the $f(R)$ model from the growth rate data assumed to be scale-independent on all the survey scales, we utilize the scale-dependent growth rate data measured in several different wavenumber bins. They are the PV measurements of $f_m\sigma_8(k,z=0)$ in five wavenumber bins, $k\in[0.005,0.02]$, $[0.02,0.05]$, $ [0.05,0.08]$, $ [0.08,0.12]$ and $[0.12,0.150]h\,\rm{Mpc^{-1}}$. We also use the RSD measurements of $f_m\sigma_8(z)$ at $z=0.25$ and $z=0.37$ in one wavenumber bin $k\in[0.005,\,0.033]h\,\rm{Mpc^{-1}}$. In each wavenumber bin, the theoretical growth rate predicted by the $f(R)$ model can be approximatively considered scale-independent. %By using the CMB+BAO+PV data, we find a peak appearing in the one-dimensional posterior distribution of the parameter $\log_{10}B_0$, showing that current PV data slightly favor a scale-dependent growth rate.
By using the CMB+BAO+PV+RSD data, we get a tight 95\% upper limit of $\log_{10}B_0<-4.1$.

We also extend our discussions to a universe with sterile neutrino species. Unlike the case in $f(R)$ gravity, the extra sterile neutrino tends to suppress the growth of structure below its free-streaming scale. We find a slight degeneracy between the $f(R)$ parameter and the sterile neutrino parameters. Due to this degeneracy, the constraint on the $f(R)$ parameter is slightly weakened to $\log_{10}B_0<-3.8$ at 2$\sigma$ level from the CMB+BAO+PV+RSD data. For the massive sterile neutrino parameters, we get $m_{\nu,{\rm{sterile}}}^{\rm{eff}}<0.62$ eV and $N_{\rm eff}<3.90$ (2$\sigma$). As a comparison, we also obtain $m_{\nu,{\rm{sterile}}}^{\rm{eff}}<0.56$ eV and $N_{\rm eff}<3.92$ (2$\sigma$) in the standard $\Lambda$CDM model.

%\newpage
\begin{acknowledgments}
%We thank Pengjie Zhang for helpful discussion.
This work was supported by the National Natural Science Foundation of
China under Grant No.~11175042, the Provincial Department of Education of
Liaoning under Grant No.~L2012087, and the Fundamental Research Funds for the
Central Universities under Grants No.~N140505002, No.~N140506002, and No.~N140504007.
\end{acknowledgments}

\end{document}